\documentclass[11pt,twoside]{article}
\usepackage{TARTU2005}
\usepackage{epsf}
\usepackage{psfig}
\usepackage{lscape}

\begin{document}
\title{Observed Metallicity Dependence of Winds from WR stars}  
\author{Paul A. Crowther} 
\affil{Dept of Physics \& Astronomy, University of Sheffield, Hounsfield Rd, Sheffield, S3 7RH,  UK}  

\begin{abstract} A review of observational evidence in favour of a
metallicity dependence of WN and WC stars is presented.  New near-IR
studies of Milky Way, LMC and SMC early-type WN stars are presented, with 
weaker
winds amongst WN stars containing hydrogen. A metallicity dependence is
supported for WN stars with hydrogen, with $dM/dt \propto Z^{\alpha}$
with $\alpha\sim0.8\pm0.2$
The influence of CNO content upon  WN subtypes is discussed. Earlier
WN spectral types are expected (and observed) at lower metallicity due to 
the 
abundance sensitivity of N\,{\sc
iii-iv} classification diagnostics.  Recent physical and
chemical results of WC stars in the Milky Way and LMC are discussed,
suggesting a metallicity dependence of $\alpha\sim 0.6\pm0.1$.  
Earlier WC spectral types are predicted (and observed) in lower 
metallicity galaxies, due to the 
dependence of the C\,{\sc iii} classification diagnostic on wind density. 
WO stars reveal lower wind velocities at lower metallicity,
whilst the situation for WN and WC stars is unclear. Finally, the
influence of a WR metallicity dependence upon the ionizing flux 
distributions and optical line luminosities is addressed, with
particular regard to I\,Zw~18.  Weaker winds at low metallicity would imply
harder ionizing flux distributions and lower line luminosities, arguing for
an substantially increased number of WR stars with respect to standard 
calibrations, exacerbating difficulties with single star 
evolutionary models at very
low metallicity.
\end{abstract}

\section{Introduction} 

Wolf-Rayet (WR) stars represent the final stages in the evolution of the
most massive stars, whose surface abundances trace the products of H
(WN subtypes) and He (WC, WO subtypes) burning. They possess fast, dense
winds, such that their spectroscopic appearance is dominated by
characteristic broad, emission lines, with photospheric lines absent.  
Indeed, narrow-band imaging surveys of nearby galaxies have identified WR
stars primarily via their emission lines (e.g. Hadfield et al. 2005), which 
can also be seen in the integrated spectra of a subset of star forming 
galaxies, known as WR galaxies.

Spectroscopically, WR stars are classified by the ratio of adjacent ions
of nitrogen (WN), carbon (WC) or oxygen (WO), with high ionization in
early subtypes, and low ionization in late subtypes (Smith, Shara \& Moffat 
1996;
Crowther, De Marco \& Barlow 1998). WN stars are also commonly subdivided 
into  strong/weak or narrow/broad emission line stars.
Observationally, earlier subtypes of both WN and WC
stars dominate at low metallicities, a tendency which has 
been attributed to increased surface C+O abundances at lower metallicity 
for WC stars (Smith \& Maeder 1991).

A metallicity dependence of the winds of massive O stars has long been
established (e.g. Garmany \& Conti 1985; Kudritzki, Pauldrach \& Puls 
1987), such that the observed population of WR stars are known to vary with 
environment (Maeder \& Meynet 1994; Massey \& Johnson 1998).  In the Milky 
Way, with high main sequence mass-loss, WR stars are relatively common with
N(WR)/N(O)$\sim$0.1 and N(WC)/N(WN)$\sim$1. In contrast, lower mass-loss
prior to the WR phase in the Small Magellanic Cloud (SMC) causes a
dramatic reduction in both the number of WR stars, N(WR)/N(O)$\sim$0.01,
and their subtype distribution, N(WC)/N(WN)$\sim$0.1.

Up until recently, the winds of WR stars were assumed to be metallicity
independent (Langer 1989). However, both observational (Crowther et al.
2002), and theoretical (Gr\"{a}fener \& Hamann 2005;  Vink \& de Koter 
2005) studies have proposed a metallicity dependence, which have been
incorporated into the latest evolutionary models (Meynet \& Maeder 2005).
Observational evidence in favour of a metallicity dependence is presented
in this review. The question of metallicity dependent winds in WR stars
has received considerable interest of late, since they represent the prime
candidates for long-duration, soft Gamma Ray Bursts (GRBs), following the
collapsar or hypernova scenario (Woosley, Eastman \& Schmidt 1999). 
Consequently, the circumstellar environment of low metallicity GRBs is 
expected to differ substantially from those in metal-rich regions 
(Eldridge et al. 2006).
%Hirschi, Meynet \& Maeder (2005) have recently suggested that WO stars are 
%the immediate precursors of GRBs. 

\section{Physical and Chemical Properties}

Prior to the development of non-LTE model atmosphere codes, 
the wind characteristics of WR stars were derived from their free-free 
excess following the technique of Wright \& Barlow (1975) via mid-IR or 
radio observations, revealing high mass-loss rates of a few 10$^{-5}
M_{\odot}$ yr$^{-1}$. Indeed, radio mass-loss rates should  be reliable 
providing the chemical composition and degree of ionization 
are appropriate to the radio forming continuum. Recent studies
claiming reduced mass-loss rates for WC9 stars (e.g. Leitherer, Chapman \& 
Koribalski 1997) are questioned by Crowther, Morris \& Smith (2006) who 
argue that helium is partially neutral in the outer stellar wind.

\begin{figure}[ht!]
\plotfiddle{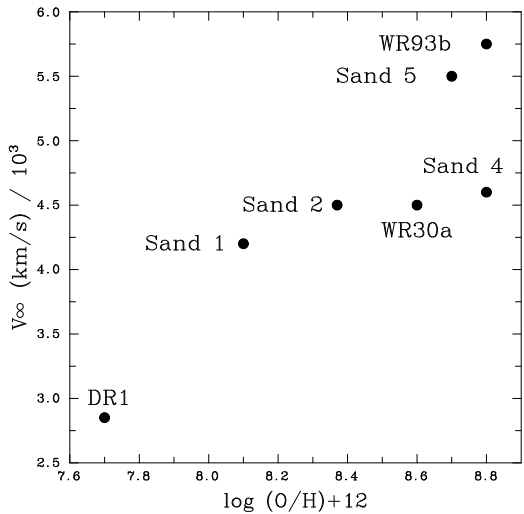}{6.0cm}{0}{50}{50}{-150}{-115}
\caption{Comparison between wind velocities of all known WO
stars from Kingsburgh et al. (1995), Kingsburgh \& Barlow 
(1995) and Drew et al. (2004). For Milky Way WO stars we 
have adopted $\log (O/H)+12$ = 8.6 for WR30a which lies 
beyond the Solar circle, 8.7 for Sand 5 which lies at the 
Solar circle, and 8.8 for Sand 4 and WR93b which lie within 
the Solar circle.}\label{velocity} \end{figure}

Terminal wind velocities, most reliably measured from UV 
P~Cygni lines (Prinja, Barlow \& Howarth 1990) range 
from $\leq$1000 km/s 
for late WN and WC subtypes to $\geq$3000 km/s for early WC stars or even
higher for WO stars (Kingsburgh, Barlow \& Storey 1995). Terminal 
velocities
can alternatively be obtained from optical emission lines, providing the
wind has reached its maximum flow rate in the line forming region. This
is generally true for stars exhibiting strong emission lines, but not 
for weak emission line stars, such as the WN stars in the SMC (Conti, 
Garmany \& Massey 1989). Differences in WR subtype distributions in 
different host galaxies 
generally hinders efforts at establishing a metallicity dependence. 
Fortunately, the presence of WO stars in a wide variety of environments, 
from the  inner Milky Way (e.g. Sand 4) to IC\,1613 (Kingsburgh \& Barlow 
1995)  permits studies of similar stars spanning an order of magnitude in 
metallicity. Fig.~\ref{velocity} compares wind velocities of all known
WO stars, and indeed suggests lower wind velocities at low metallicity,
despite small number statistics.

Hydrogen abundances in WN stars may be derived from the Pickering-Balmer
series (Conti, Leep \& Perry 1983) and support severe H depletion, whilst 
high carbon abundances of C/He$\geq$0.1 can be determined for early WC 
stars 
from recombination line theory (Smith \& Hummer 1988).

The development of non-LTE model atmospheres by D.J.~Hillier and
W.-R.~Hamann and coworkers during the 1980s and 1990s has permitted great
progress via the determination of physical and chemical properties by
spectral line fitting, via diagnostic lines of He and/or N in WN stars and
He and/or C in WC stars. The latest versions of these codes typically treat
non-LTE in a spherical, expanding, extended atmosphere considering the
effects of metal line blanketing (CMFGEN: Hillier \& Miller 1998; PoWR:
Gr\"{a}fener \& Hamann 2005). Historically, O stars were analysed
using plane-parallel codes (e.g.  TLUSTY: Hubeny \& Lanz 1995), whilst 
current models (e.g. FASTWIND, Puls et al. 2005)  now also account for 
winds and line blanketing.

Clumping is incorporated into these codes following an approximate manner
via a volume filling factor, $f$, such that identical recombination line
profiles -- scaling with the square of the density -- result if
$\dot{M}/\sqrt(f)$ is held constant. The primary diagnostics constraining
clumping are the electron scattering wings, with a linear dependence upon
density (Hillier 1991). The assumption of spherical symmetry appears to be 
reasonable for the majority of WR stars, since spectropolarimetric studies 
indicate only 5 out of 29 WR stars show any persistent line effect 
(Harries, Howarth \&  Hillier  
1998). Stars which do reveal significant equator-to-pole density ratios of
2--3 tend to possess strong winds and  span late WN (with hydrogen), early 
WN (no hydrogen) and WC subtypes.

%(albeit a binary system).

\section{Metallicity dependent WN winds?}

%\subsection{Metallicity dependence?}

Smith \& Willis (1983) compared the properties of Large Magellanic Cloud
(LMC) to Milky Way WR stars and concluded there was no significant
differences between the two populations. These conclusions were supported
by Koesterke et al.  (1991) and later Hamann \& Koesterke (2000) from
detailed non-LTE modelling although a large scatter in mass-loss rates
within each parent galaxy was revealed. One might conclude that there was
no metallicity dependence, or that any differences are too subtle to be
identified from the narrow metallicity range spanned by the Milky Way and
LMC, given the multiple evolutionary channels available to stars entering
the WN stage.

\begin{figure}[htb]
\plotfiddle{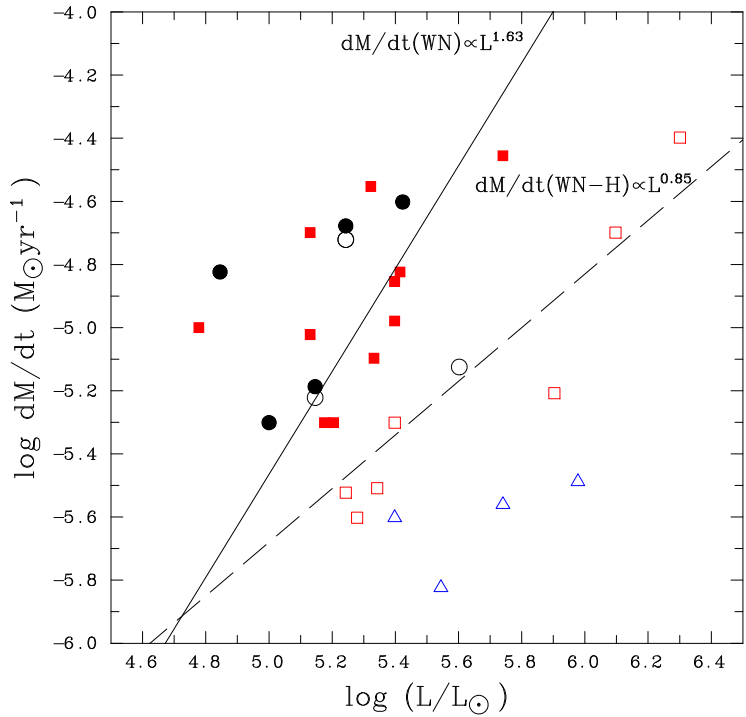}{9.5cm}{0}{70}{70}{-200}{-150}
\caption{Comparison between mass-loss rates and luminosities of 
Milky Way (circles), LMC (squares) and SMC (triangle) WN3--6 stars
analysed using IRTF/SpeX (Milky Way) and NTT/SofI (Magellanic Clouds)
near-IR spectroscopy. Filled (open) symbols correspond to stars without 
(with) surface hydrogen. We  include the Nugis \& Lamers (2000) luminosity 
dependence 
(solid line) plus the fit to the Milky Way and LMC WN stars
containing hydrogen, excluding HD\,192163 (dashed line). The 
SMC stars which also contain hydrogen lie --0.4 dex below this 
fit.}\label{nearir} \end{figure}

Crowther (2000) analysed the sole (at that time) single WN star in the
SMC, Sk~41 (WN6ha). In comparison to LMC and Galactic late-type WN
counterparts, Sk~41 was revealed to possess a low wind velocity and
mass-loss rate. A larger sample of single low metallicity WN stars was
needed for definitive conclusions regarding a metallicity dependence. This
was provided by Foellmi, Moffat \& Guerrero (2003a) who concluded that many 
SMC WN stars which were hitherto considered to be binaries were apparently
single. Observationally, SMC WN stars are well known to possess narrow,
weak emission lines with respect to higher metallicity counterparts (Conti 
et al. 1989).

%\subsection{Near-IR studies of early WN stars}

The strength of He\,{\sc ii} $\lambda$4686 is related both to the wind
density and ionization, whilst He\,{\sc i} 1.083$\mu$m is the primary wind
density diagnostic in the optical/near-IR (Howarth \& Schmutz 1992). In
order to quantify the winds from weak-lined, single SMC stars with respect
to Milky Way and LMC counterparts we have obtained near-IR spectroscopy
from NTT/SofI for 4 SMC and 18 LMC WN3--6 stars from the samples of
Foellmi et al. (2003ab) in collaboration with C. Foellmi and W. Vacca. In
addition, near-IR spectroscopy of 8 Milky Way WN3--6 stars at known
distances have been obtained by W. Vacca from IRTF/SpeX.

We present CMFGEN results from our near-IR studies of WN stars 
in 
Fig.~\ref{nearir}, where we have distinguished between stars containing
surface hydrogen  and those that do not. Luminosities (in 
$L_{\odot}$) range from  10$^{4.8}$ to 10$^{6.3}$ whilst mass-loss rates
span 10$^{-5.9}$ to 10$^{-4.4} M_{\odot}$ yr$^{-1}$.
It is apparent that the
stars without hydrogen possess the strongest winds, with the exception of 
HD~192163 (Crowther \& Smith 1996). The generic WN mass-loss luminosity 
relation of Nugis \& Lamers (2000) for zero hydrogen content is 
indicated with exponent dM/dt $\propto L^{1.6}$, whilst a fit to our
sample would suggest a softer dependence of dM/dt $\propto L^{0.6}$
suggesting less extreme mass-loss rates at the highest luminosities.

Since the number of Milky Way and LMC stars with atmospheric hydrogen
is fairly small, we have fit the combined sample (excluding HD~192163), 
suggesting a dependence
of \[ \log (\dot{M}/M_{\odot}{\rm yr}^{-1})  = 0.85 \log (L/L_{\odot}) - 
9.93.\]
It is apparent that the SMC stars possess even weaker winds. Assuming
a similar luminosity dependence, one would require a --0.4 dex offset
for the SMC stars (1/5 $Z_{\odot}$) with respect to the Milky 
Way/LMC stars (1/2 to 1 $Z_{\odot}$), i.e. a metallicity dependence of
dM/dt $\propto Z_{\odot}^{\alpha}$ with $\alpha \sim 0.8\pm0.2$. Note that 
negligible He\,{\sc i} emission was observed for two SMC stars, such
that an effective temperature of $\sim$85kK was adopted on the basis of
the observed optical nitrogen spectrum, with the mass-loss rate resulting
from He\,{\sc ii} $\lambda$4686 emission.

%\subsection{Subtype dependence}

\section{Metallicity dependent WC winds?}

%\subsection{Metallicity dependence?}

Gr\"afener et al. (1998) provided the first modern quantitative
comparison between LMC and Milky Way WC stars, suggesting either 
a dependence of dM/dt $\propto L^{0.75}$ for the combined sample,
or a steeper dependence for the Milky Way WC5--8 stars with
dM/dt $\propto L^{1.5}$ with weaker winds for the LMC WC4 stars.
Comparisons at still lower metallicity are hindered because the sole 
carbon-sequence member of the SMC is the WO binary Sand~1 (Sk~188).

\begin{figure}[htb]
\plotfiddle{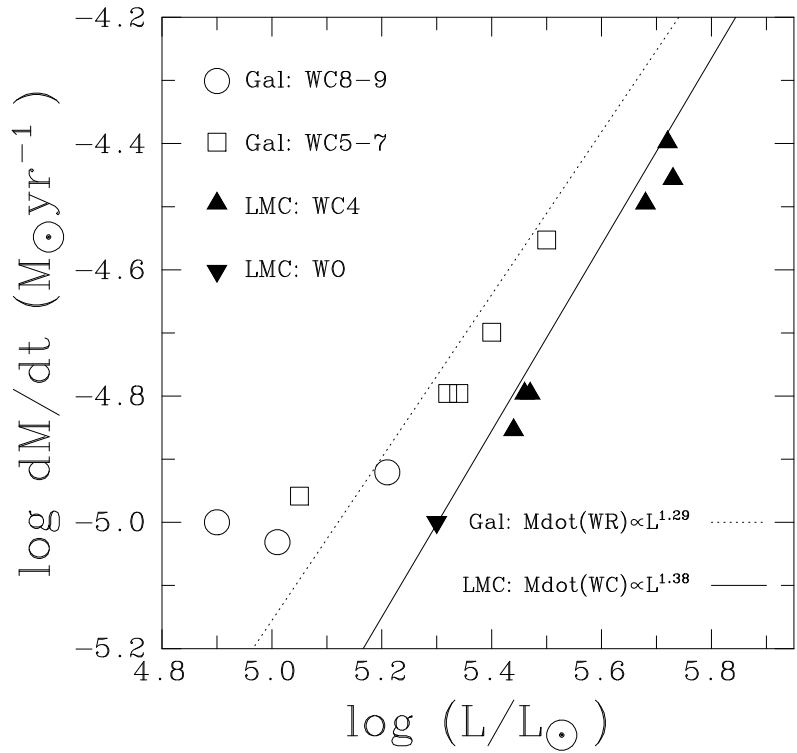}{9.5cm}{0}{70}{70}{-200}{-150}
\caption{Comparison between mass-loss rates and luminosities of 
Milky Way (open) and LMC (filled) WC and WO stars, taken
from Crowther et al. (2002, 2005). We  include the 
Nugis \& Lamers (2000) luminosity dependence (dotted line) 
for WR stars with composition  C/He=0.2 and C/O=4 by number, plus the fit 
(solid line) to LMC stars by Crowther et al. (2002).}\label{wc} 
\end{figure}

Subsequently, Crowther et al. (2002) re-analysed the LMC WC4 sample
of  Gr\"afener et al. (1998) based on line blanketed models together
with an increased sample of Milky Way stars which has been 
analysed in the  same manner.
Fig.~\ref{wc} presents a comparison between the mass-loss rates and
luminosities of Milky Way to LMC stars from Crowther et al. (2002), 
including HD~164270 (WC9) from Crowther et al. (2006). The early 
WC Milky Way  stars closely follow the Nugis \& Lamers (2000) generic
calibration, assuming C/He=0.2 and C/O=4 by number, whilst a 
fit to the LMC sample suggests a dependence 
of 
\[ \log (\dot{M}/M_{\odot}{\rm yr}^{-1})  = 1.38 \log (L/L_{\odot}) - 
12.35\]
i.e. revealing a similar slope to Nugis \& Lamers (2000), albeit offset by 
--0.2 dex. Consequently, results for WC stars close to Solar metallicity 
suggest a $Z$ dependence with dM/dt $\propto Z^{\alpha}$, with 
$\alpha\sim0.6 \pm 0.1$.

\section{Metallicity dependence of WR subtypes}

%\subsection{WN stars}

Empirically, the majority of WN stars in the Milky Way are late-type
with N(WNL):N(WNE)$\sim$3:2 whilst those in the LMC and especially the SMC 
are early-type, with N(WNL):N(WNE)$\sim$1:5. Within the Milky Way most 
late-type stars contain hydrogen and most early-type
stars do not, which is no longer true in the Magellanic Clouds. There are
two atmospheric factors contributing towards the earlier subtypes at 
lower  metallicities, namely the fixed CNO at a particular metallicity and 
the apparent decrease in wind strength with metallicity. 

CNO compromises $\sim$1.1\% by mass of the Solar photosphere (Asplund et 
al. 2004) versus 0.48\% in the LMC and 0.24\% in the SMC (Russell \& Dopita
1990). Since WN stars typically exhibit CNO equilibrium abundances, there 
is a clearly maximum nitrogen content available within a particular 
environment. Crowther (2000) demonstrated that for otherwise
identical parameters, regardless of metallicity dependent mass-loss 
rates, a decreased nitrogen content at lower metallicity favours an earlier
subtype, due to the strong abundance sensitivity of N\,{\sc 
iii} $\lambda$4634--41 and weak sensitivity of N\,{\sc iv} $\lambda$4058.

If WN winds do scale with metallicity, one also expects earlier
subtypes since high wind densities at high metallicities will efficiently
cool the wind through metal lines (e.g. Hillier 1989), causing 
recombination from high ionization stages (e.g. N$^{5+}$) to lower
ions (e.g. N$^{3+}$) close to the optical line formation region, which
does not occur for low wind densities. Consequently, both factors favour 
late subtypes at high metallicity, and early subtypes at low metallicity,
as generally observed.

%\subsection{WC stars}

The observational trend towards earlier  WC subtypes at lower metallicity,
together with early recombination line studies of early WC stars, 
led Smith \& Maeder (1991) to suggest
that early WC stars are more carbon-rich than late WC stars. Koesterke \& 
Hamann (1995) analysed a large sample of Galactic WC5--8 stars, but did 
not confirm a subtype dependence. Crowther et al. (2002) supported the
conclusions of Koesterke \& Hamann, since the range of (C+O)/He abundances 
in LMC WC4 stars were similar to those of Milky Way WC5--8 stars.

Crowther et al. (2002) argued that weaker winds for early WC subtypes was
the prime reason for a trend towards early subtypes at low metallicities.
Indeed, their fig.~12 illustrated that at fixed stellar parameters and
chemical composition, a reduction in mass-loss rate by only a factor of
two causes a WC7 star to become a WC4 subtype. The cause of this dramatic
shift is due primarily to the sensitivity of the classification line
C\,{\sc iii} $\lambda$5696 to mass-loss. Other C\,{\sc iii} lines,
such as C\,{\sc iii} $\lambda$6740, are relatively insensitive to 
such modest changes in wind density.
If early WC stars dominate at low metallicity, as is the case for the
LMC, SMC and IC\,10 (Crowther et al. 2003),
%
%\footnote{Curiously, one of 
% fourteen WC stars in IC\,10, \#10 from Massey et al. (1992) has a WC7 
% subtype},  
%
one would expect that late WC stars dominate at high 
metallicity. Indeed, 
all 5 WR stars  recently identified in the inner Milky Way by Hopewell et 
al. (2005)  were of WC9 subtype. Hadfield et al. (2005) also found that 
WC8--9 subtypes dominate the WC population of the  metal-rich
spiral galaxy M~83.

\begin{figure}[htb]
\plotfiddle{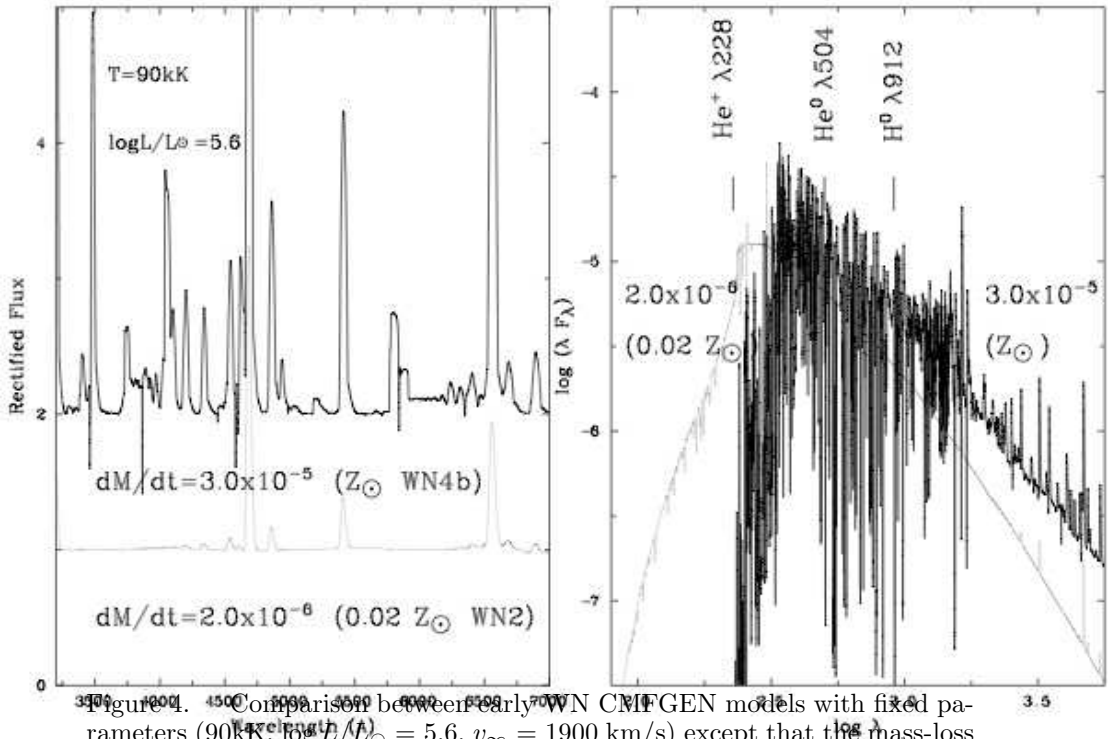}{8.5cm}{-90}{52.5}{52.5}{-210}{260}
\caption{Comparison between early WN CMFGEN models with fixed parameters
(90kK, $\log L/L_{\odot}$ = 5.6, $v_{\infty} $ = 1900 km/s) except
that the mass-loss rates (and metal abundances) differ by a factor of
15 (50). The high mass-loss rate model closely matches the spectrum
of HD~50896 (WN4b, Morris et al. 2004) whilst the low mass-loss rate
model might be representative of an early WN star in I\,Zw~18 (Crowther 
\& Hadfield 2006).}
\label{ionize} \end{figure}

\section{Impact of weak WR winds at lower metallicity: I\,Zw~18}

To date, several studies have identified WN and WC stars in the 
very metal poor (1/50 $Z_{\odot}$) galaxy I\,Zw~18. Izotov et  al. 
estimated a  total of 17 late WN stars plus 5 early WC stars, whilst WC 
populations were identified by Legrand et al. (1997) and Brown et al. 
(2002). This galaxy lies at 10--15\,Mpc, so one has to infer 
individual properties from extrapolation of nearby, resolved WR 
populations.

Schmutz, Leitherer \& Gruenwald (1992) stressed the importance of stellar 
wind density on the ionizing flux distributions of WR stars, such that 
emission at
energies above the He$^{+}$ edge ($\lambda < 228$\AA) relies upon
the WR wind being relatively transparent. High metallicity, strong winds
will generally produce strong optical emission lines, though negligible 
hard ionizing radiation, while low metallicity, weak winds will conversely 
produce weak optical emission lines and prodigious hard ionizing 
radiation (Smith, Norris \& Crowther 2002). Therefore, if low metallicity 
WR stars -- such as those observed in I\,Zw~18 --  do possess weak winds, 
they  would be expected to possess (difficult to detect) weak emission 
lines, faint optical  continuua and hard extreme UV radiation. Indeed, 
strong nebular He\,{\sc  ii} $\lambda$4686 is observed in I\,Zw~18 whose 
origin may be due to WR stars.

To illustrate the effect of reduced wind density upon the ionizing flux 
distribution of WN models, Fig.~\ref{ionize} compares the rectified 
optical spectra and ionizing flux distributions for the line
blanketed CMFGEN model of the strong-lined Galactic WN4b star HD~50896 
(Morris, Crowther \& Houck   2004) with an identical model, except that the 
elemental 
abundances  have been reduced by a factor of 50, to mimic the abundances 
of I\,Zw~18,  with the mass-loss rate reduced by a factor of 
50$^{0.7}\sim15$ from  3$\times 10^{-5}$ to 2$\times 10^{-6} M_{\odot}$ 
yr$^{-1}$. It is apparent that the low metallicity WR star has a 
much harder ionizing flux distribution, with significant emission shortward
of the He$^{+}$ edge at 228\AA. In addition, the weak wind model
has a factor of 
4 times smaller He\,{\sc ii} $\lambda$4686 equivalent width, plus 
a factor of 5 times weaker optical continuum, such that the He\,{\sc ii} 
$\lambda$4686 
line  luminosity is  reduced by a factor of 20 relative to the Solar 
counterpart.

\begin{figure}[htb]
\plotfiddle{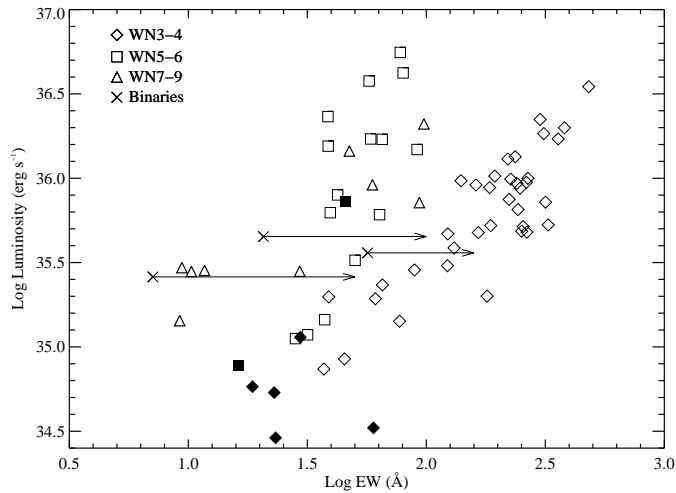}{6.0cm}{90}{39}{39}{150}{-30}
\caption{Comparison between the He\,{\sc ii} 
$\lambda$4686 line luminosities and equivalent widths
for single LMC (open) and SMC (filled) early, mid and late WN stars 
from Crowther \& Hadfield (2006). We additionally include 3 SMC WN3-4 
binaries (crosses), in which arrows indicate equivalent widths corrected
for dilution by the O-type companion. The average LMC line luminosities 
agree well with Schaerer \& Vacca (1998) despite the 
large scatter, whilst the SMC values lie a factor of 
4--5 times lower (including binaries), with potentially large consequences 
for the absolute determination of WR populations at low
metallicity.}\label{luminosity} 
\end{figure}

At present, average Milky Way/LMC WN and WC line luminosities are used to
determine the number of WR stars in unresolved star forming regions at
{\it all} metallicities (Schaerer \& Vacca 1998). If individual WR line
luminosities are lower at lower metallicity, this would imply a larger
number of WR stars than is presently assumed.  Empirically, 
Fig.~\ref{luminosity} illustrates  that indeed WN stars in the SMC 
do possess lower He\,{\sc ii} $\lambda$4686 line luminosities than their 
counterparts in the LMC, by factors of 5 (early WN) or 4 (mid WN) on 
average (Crowther \& Hadfield 2006).

De Mello et al. (1998)  argued that allowing for the WC contribution 
to the blue WR feature in the observations  of Izotov et al. 
(1997) would reduce the WN content to  $\sim$4 late-type stars, based on 
the  standard line luminosity calibrations.  If the WN stars observed in 
I\,Zw~18 are early WN stars, which are more common at lower metallicity,
one would need  to increase that total 
by a factor of up to $\sim$30, to $\sim$120 for their assumed distance. 
De  Mello et  al. (1998)   claimed  reasonable agreement between  
instantaneous burst  models at    1/50$Z_{\odot}$ with observations. This 
agreement would naturally be lost  if the  absolute  number of WR stars 
was increased by such a large factor.

\begin{figure}[htb]
\plotfiddle{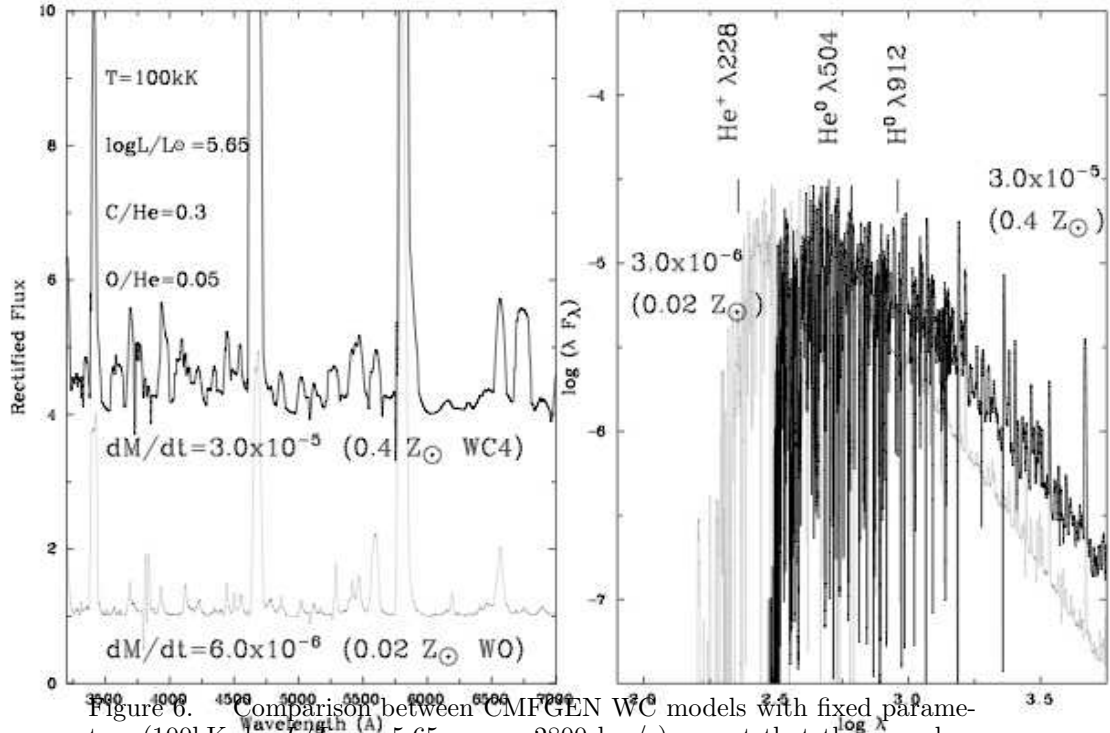}{8.5cm}{-90}{52.5}{52.5}{-210}{260}
\caption{Comparison between CMFGEN WC models with fixed parameters
(100kK, $\log L/L_{\odot}$ = 5.65, $v_{\infty} $ = 2800 km/s) except
that the mass-loss rates differ by a factor of 5, and abundances for 
elements heavier than neon differ by a factor of 20. The high mass-loss 
rate model  closely  matches the spectrum of the LMC WC4 star HD~37026 
(Crowther et al. 2002) whilst  the low mass-loss rate model might be 
representative of a WO4 star in I\,Zw~18 (Crowther \& Hadfield 2006).}
\label{wo} \end{figure}

How would the properties  of low metallicity, weak wind WC stars differ from 
those in the Milky Way? Fig.~\ref{wo} compares a CMFGEN model for the LMC WC4 
star HD~37026  from Crowther et al. (2002) with an identical model, except 
that the  heavy element abundances (beyond Ne) have  been reduced from 
$\sim$1/2 $Z_{\odot}$  to 1/50  $Z_{\odot}$ (I\,Zw~18),  and
the model mass-loss rate has been  reduced by a  factor of  
25$^{0.5}\sim 5$ from 3$\times 10^{-5}$ to  6$\times 10^{-6}$ 
$M_{\odot}$yr$^{-1}$. As with the WN case, the low mass-loss model 
displays an earlier spectral type (WO), and has a much harder ionizing 
spectrum, including significant emission shortward of the He$^{+}$ edge
at 228\AA. Indeed, some WO stars are associated with H\,{\sc ii} 
regions containing nebular He\,{\sc ii} $\lambda$4686 
(e.g. DR1, Kingsburgh \& Barlow 1995).

Observationally, WO stars, showing strong O\,{\sc vi-vi} emission dominate
at the lowest metallicities (SMC, IC\,1613).  Kingsburgh et al. (1995)  
demonstrated that WO stars exhibit high (C+O)/He abundances, which was
confirmed by detailed modelling of the single LMC WO star Sand~2 by
Crowther et al.  (2000). Why are WO stars preferentially observed in
low metallicity regions?  Strong, carbon-rich winds very effectively
recombine high ionization stages, such as O$^{6+}$, to lower ionization,
e.g.  O$^{5+}$ or O$^{4+}$, in the optical line formation region.
One would therefore expect a 
dominant weak-lined WC or WO  population in I\,Zw~18. Izotov et al. (1997) 
and Legrand et  al. (1997) both identified broad 
(FWHM=50$\pm$10\AA) C\,{\sc iv}
$\lambda\lambda$5801--12 emission in I\,Zw~18 which they attributed to WC
stars, whilst Brown et al. (2002)  identified broad C\,{\sc iv}
$\lambda\lambda$1548--51 emission, which they also attributed to WC stars.
Neither of these observations preclude a WO origin,
since their line widths are known to decrease in lower metallicity
regions (recall Fig.~\ref{velocity}). Indeed, the lowest known metallicity WO 
star,   DR1 in  IC~1613, has FWHM(C\,{\sc iv} $\lambda\lambda$5801--12) 
$\sim$70\AA. 

\begin{figure}[htb]
\plotfiddle{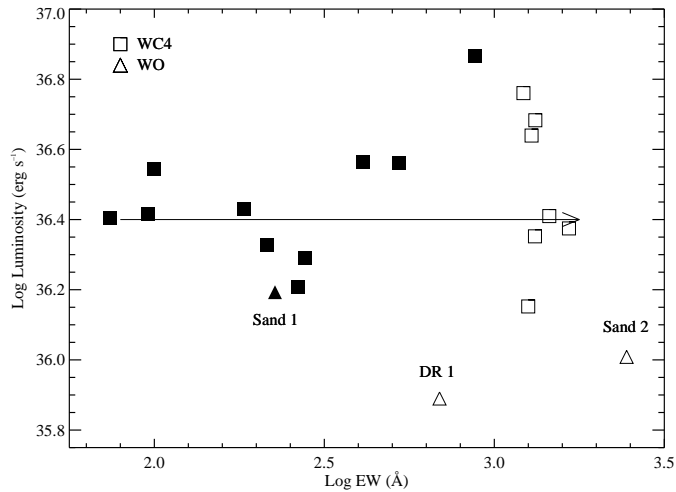}{6.0cm}{90}{39}{39}{150}{-30}
\caption{Comparison between the C\,{\sc iv} 
$\lambda$5801--12 line luminosities and equivalent widths for
single (open) and binary (filled) WC4 and WO stars at known distance (Crowther \& Hadfield 2006). 
All the WC4 stars are LMC members, whilst
the WO stars are Sand~1 (SMC), Sand~2 (LMC) and Dr~1 (IC\,1613)
from Kingsburgh et al. (1995) and Kingsburgh \& Barlow (1995).
An estimate of line dilution is indicated for BAT99-10 (WC4+O).}
\label{wc_luminosity} 
\end{figure}

We compare the observed C\,{\sc iv} $\lambda\lambda$5801--12 line
luminosities of single and binary WC4 and WO stars in Fig.~\ref{wc_luminosity}, 
revealing a factor of $\sim$3 times lower luminosity for WO stars. Indeed, the two 
WC models presented in Fig.~\ref{wo} differ by a factor of 6 in C\,{\sc iv}
$\lambda$5801--12 line luminosity. Consequently, if the I\,Zw~18 carbon
sequence WR stars are closer analogues of the WO stars, the number of stars
would again need to be increased upwards, from $\sim$5 to $\sim$30
based on the observations of Izotov et al. (1997).

In summary, a natural consequence of metallicity 
dependent WR winds would be that the number of unresolved  WR stars at 
low metallicities are presently severely underestimated by application 
of calibrations appropriate to LMC/Milky Way WR stars (Schaerer \& Vacca 
1998). If high spatial  resolution observations of WR  stars in I\,Zw~18 
reveal properties similar to Milky Way counterparts one would have to 
question their dependence on metallicity proposed here. Finally, reduced wind 
strengths from WR stars at low metallicities impacts  upon the immediate 
circumstellar  environment of long duration GRB afterglows, particularly 
since the host galaxies of high-redshift GRBs tend to be metal-poor (e.g.
Chen et al. 2005).

%\section{Summary}

\acknowledgements The near-IR studies of WN stars presented here are in 
collaboration with Bill Vacca, Cedric Foellmi and Lucy Hadfield.

\end{document}